\documentclass[structabstract]{aa}  
\usepackage{graphicx}
\usepackage{txfonts}
\begin{document}
   \title{On the Magnetic Field Strength of Active Region Filaments}



   \author{C. Kuckein\inst{1}
   \and R. Centeno\inst{2}
   \and V. Mart\'inez Pillet\inst{1}
   \and R. Casini\inst{2}
   \and R. Manso Sainz\inst{1}
   \and T. Shimizu\inst{3}}

  \institute{Instituto de Astrof\'\i sica de Canarias, V\'\i a 
L\'{a}ctea s/n, E-38205 La Laguna, Tenerife, Spain\\
   \email{ckuckein@iac.es}
  \and High Altitude Observatory (NCAR), Boulder, CO 80301
  \and Institute of Space and Astronautical Science, JAXA, Sagamihara, 
Kanagawa 229-8510, Japan}

  \date{Received date / Accepted date}

 
  \abstract
   {}
   {We study the vector magnetic field of a filament observed over a compact Active
Region Neutral Line.}
   {Spectropolarimetric data acquired with TIP-II (VTT, Tenerife, Spain) of the 10830 \AA\ 
spectral region provide full Stokes vectors which were analyzed using three different methods: 
magnetograph analysis, Milne-Eddington inversions and PCA-based atomic polarization inversions.}
   {The inferred magnetic field strengths in the filament are of the order of 600 - 700 G by
all these three methods. Longitudinal fields are found in the range of 100 - 200 G whereas the 
transverse components become dominant, with fields as large as 500 - 600 G. 
We find strong transverse fields near the Neutral Line also at photospheric levels.}
   {Our analysis indicates that strong (higher than 500 G, but below kG) 
transverse magnetic fields are present in Active Region filaments. This corresponds to the highest field
strengths reliably measured in these structures. The profiles of the Helium 10830 \AA~
lines observed in this Active Region filament are dominated by the Zeeman effect.}

   \authorrunning{Kuckein, Centeno, Mart\'\i nez Pillet et al.}
   \titlerunning{Magnetic Field Strength of Active Region Filaments}
   \maketitle

\section{Introduction}

The magnetic field strength of Active Region (AR) filaments has long
remained poorly known or understood.  The situation for quiescent 
filaments is notably more satisfactory since the early 
measurements back in the 70s (see, e.g., the review by 
\cite{ariste&aulanier07}). Field strengths measured in quiescent
structures, mostly using the Hanle effect on the linear polarization
of He {\sc i} D$_3$ at 5876\,\AA\ (e.g., \cite{SB77}), were found 
to be in the 
range of 3--15\,G (\cite{leroy83}; see also the review by 
\cite{anzer&heinzel07}). However, more recent measurements, which also took 
into account circular polarization, consistently show a tendency 
towards larger field strengths. For example, the He {\sc i} 10830 
\AA~investigation of \cite{trujillo02} and the He {\sc i} D$_3$ of 
\cite{casini03} found, respectively, field strengths of the order of 20--40\,G and 
10--20\,G (with field strengths up to 80\,G in the latter set of observations). With 
similar techniques, using carefully inverted observations of 
the \mbox{He {\sc i}} 10830\,\AA\ lines, \cite{merenda06} found field strengths of 
about 30\,G. Other recent works that also obtained relatively strong filament 
magnetic fields as compared with older measurements are \cite{paletou01} 
and \cite{wiehr&bianda03}.
Of particular interest are the high field strength areas revealed by 
Zeeman-dominated Stokes $V$ profiles in conjunction with scattering 
dominated $Q$ and $U$ linear polarization signals obtained by 
\cite{casini03}, showing how important it is to use all four Stokes
parameters for a proper determination of the field strength in prominences.

Well established properties of AR filaments are their systematically 
stronger field strengths as compared to their quiescent counterparts and that 
they lie lower in the atmosphere (see, e.g., \cite{aulanier&demoulin03}).  
Zeeman observations, using coronagraphs, provided field strengths for 
AR prominences in the range of 50--200\,G
(\cite{tandberghanssen&mckimmalville74}, and references therein).
However, the intrinsic difficulties of observing such low-lying structures
near the limb and the use of H$\alpha$ magnetography render these results somewhat
questionable (see \cite{lopezariste05} for the problems associated with using a 
Zeeman-based formulation of the measurements made in H$\alpha$). 
Perhaps, the most relevant modern estimate of the field strength in 
AR prominences was provided by \cite{wiehr&stellmacher91}, who
measured the longitudinal magnetic field using the Stokes $I$ and 
$V$ profiles of the \mbox{Ca {\sc ii}} IR triplet, finding a value of 150\,G, 
compatible with previous measurements. 
It is important for the discussion presented in our paper
to note that, in that observational study, linear polarization 
signals ($Q$ and $U$) were not included. The only measurements of strong magnetic fields using \mbox{He {\sc i}} 10830\,\AA\ full Stokes vector that we are aware of, belong to a multicomponent flaring active region and reached values of $\approx$ 380\,G (see \cite{sasso07}).

Because measurements are scarce, we are still far from having a 
complete picture of the field strengths that permeate solar filaments.
Some progress has come from using photospheric distributions of (mostly
longitudinal) fields that are extrapolated into the corona (usually using models of
constant-$\alpha$ force-free field). The study by Aulanier \& D{\'e}moulin (2003) is of
particular relevance to the present work. According to these authors, the 
field strength of filaments found from the extrapolations is about 3\,G 
for the quiescent case, 40\,G for active filaments (called ``plage
case'' in their work) and 15\,G for intermediate cases. 
The typical height in the atmosphere of a filament base ranges from 20 to 10\,Mm, or 
even lower, as one moves from quiescent to AR structures. These authors also 
find positive field gradients with height that result in values of 
0.1--1\,G\,Mm$^{-1}$, i.e., stronger fields are higher up in the atmosphere.
If these gradients are used to extrapolate down into the photosphere, the 
fields there would be very similar to those found some Mm up in the 
corona. For example a 10\,Mm-height active filament with 100\,G in the 
corona would have no less than 90\,G close to the photosphere. The
argument could also be turned around to start from the fields measured 
near the photosphere, close to AR Neutral Lines (NLs), and 
infer what the fields could be high in the corona. In the case of kG-strong 
plage fields, with high density areal filling factors (that can 
reach up to 50\%; \cite{martinezpillet97}), one could expect fields 
of several hundred Gauss within the filaments. Indeed, 
Aulanier \& D{\'e}moulin (1998, 2003) provide arguments supporting a 
relationship between the photospheric fields at
the base of the filament and the filament fields up in the corona. They  
conclude that the stronger the photospheric fields at the 
base of the filament, the stronger the field in the filament itself. 
Similarly, the separation of the two opposite polarity regions
scales inversely with the field strength of the filament, the latter being larger
whenever the two polarities are closer together. Dense (highly packed)
fields in the photosphere correspond indeed to 
the photospheric configuration found below AR filaments as observed 
recently by \cite{lites05} and \cite{okamoto08}. In both 
cases, they find opposite polarity ``abutted'' plage fields at the 
NL, with sheared  horizontal fields in the hecto-Gauss range 
and relatively high filling factors. These abutted field configurations seem to 
also correspond with low-lying filaments structures. \cite{lites05}
comments that the height of the filaments on top of the abutted plage 
fields is no more than 2.5\,Mm. Thus, high density (i.e., large 
filling factor) horizontal plage fields near AR NLs, together with the 
inferences from theoretical modeling (\cite{aulanier&demoulin98}, 2003),
 would indicate that fields of several hundred
Gauss can be expected in low-lying AR filaments. 

Given the observation that energetic Coronal Mass Ejections are often associated with
AR filament eruptions (see, e.g., \cite{manchester08} and \cite{low01} for
a review), it is highly
desirable to develop diagnostic tools for direct measurements of
the AR filament magnetic field and its evolution, from its emergence 
(see Okamoto et al. 2008) to the erupting phase. In this work, we present a 
clear diagnostic tool of how this can be achieved using full Stokes polarimetry 
of the \mbox{He {\sc i}} 10830\,\AA\ lines.
\begin{figure*}
\centering
\includegraphics[width=6cm,height=5.9cm]{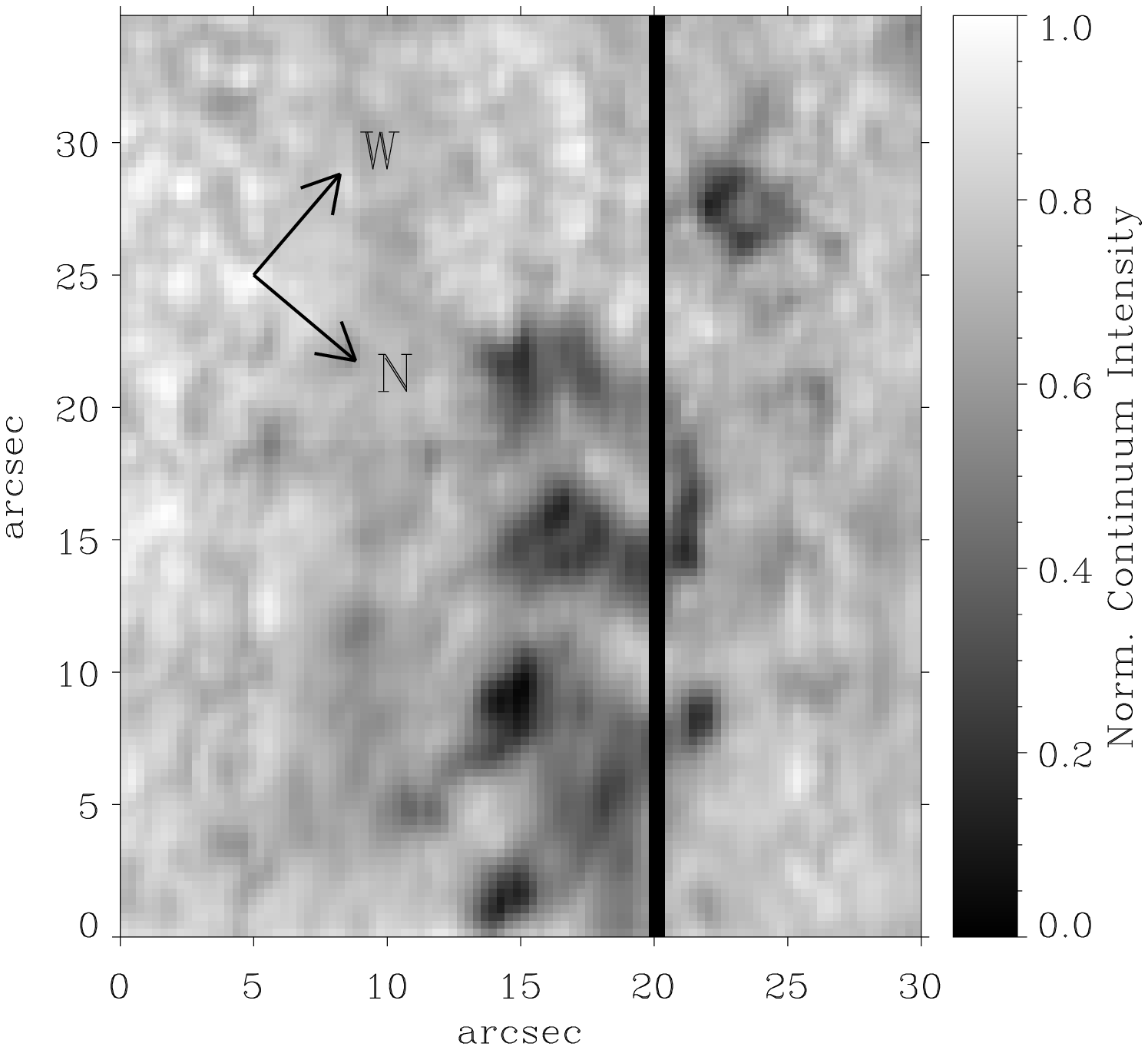}
\includegraphics[width=6cm,height=5.9cm]{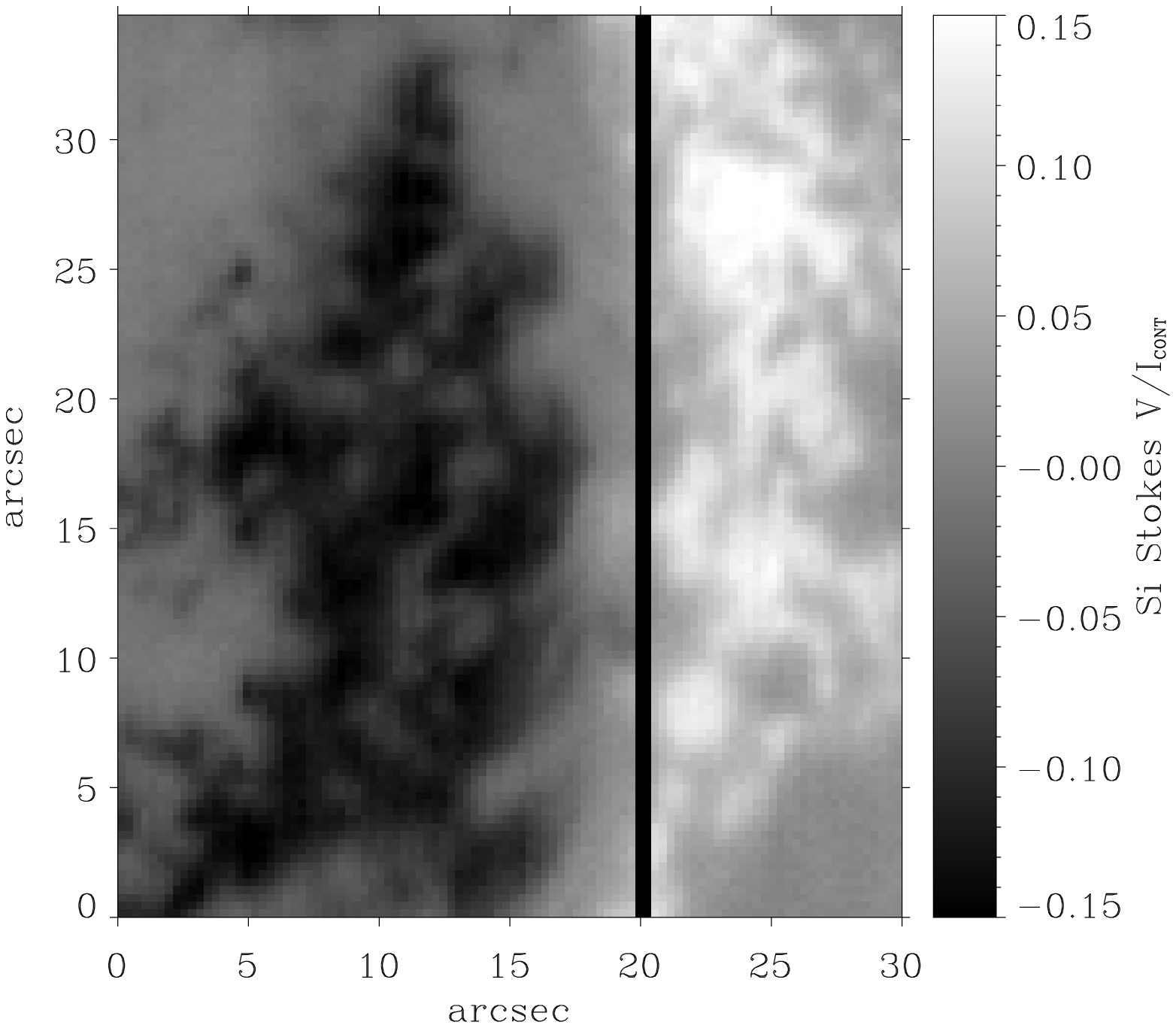}
\\
\includegraphics[width=0.49\linewidth]{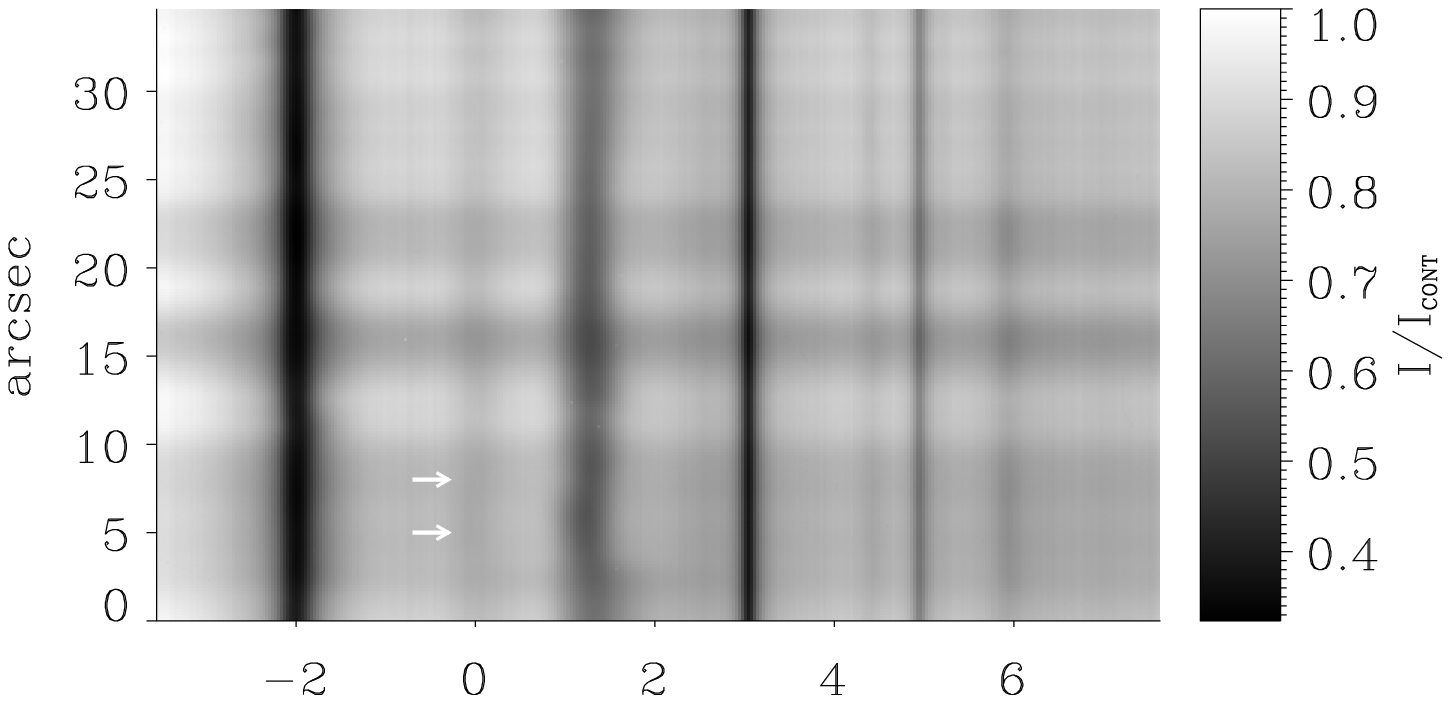}
\includegraphics[width=0.49\linewidth]{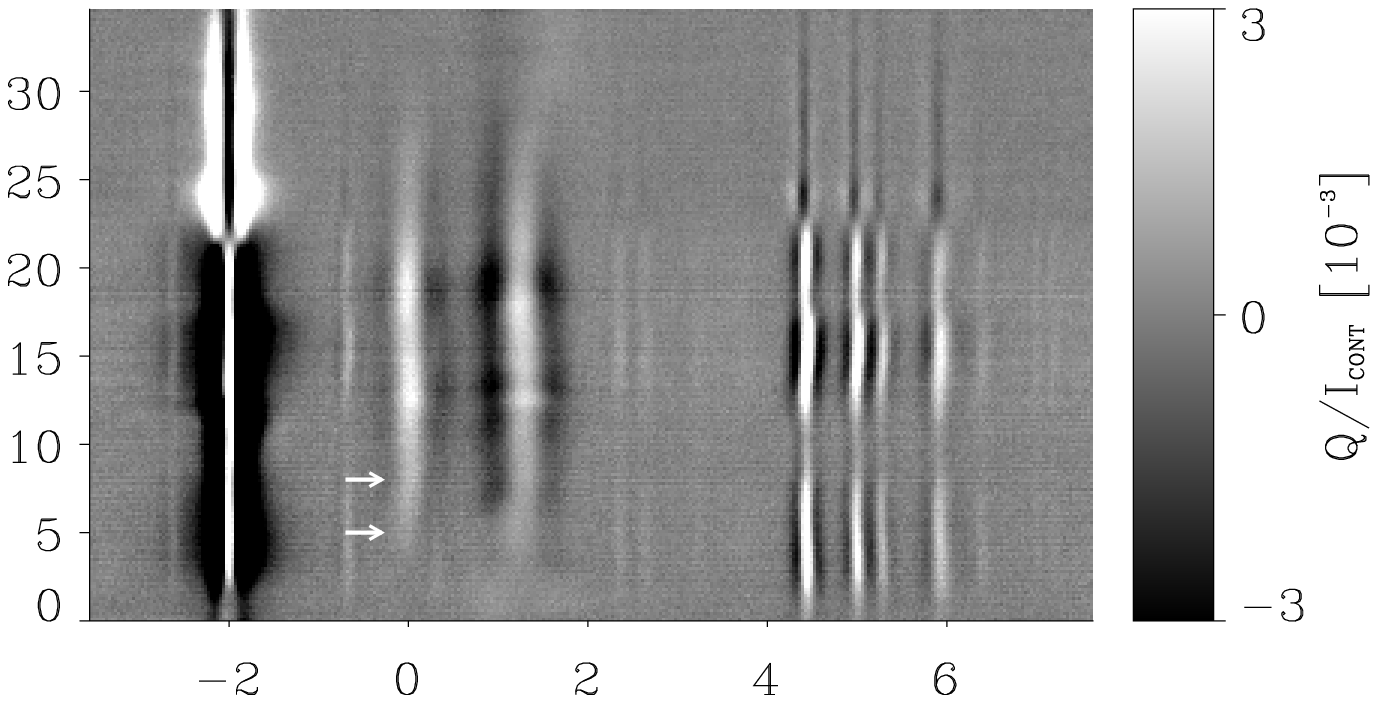}
\includegraphics[width=0.49\linewidth]{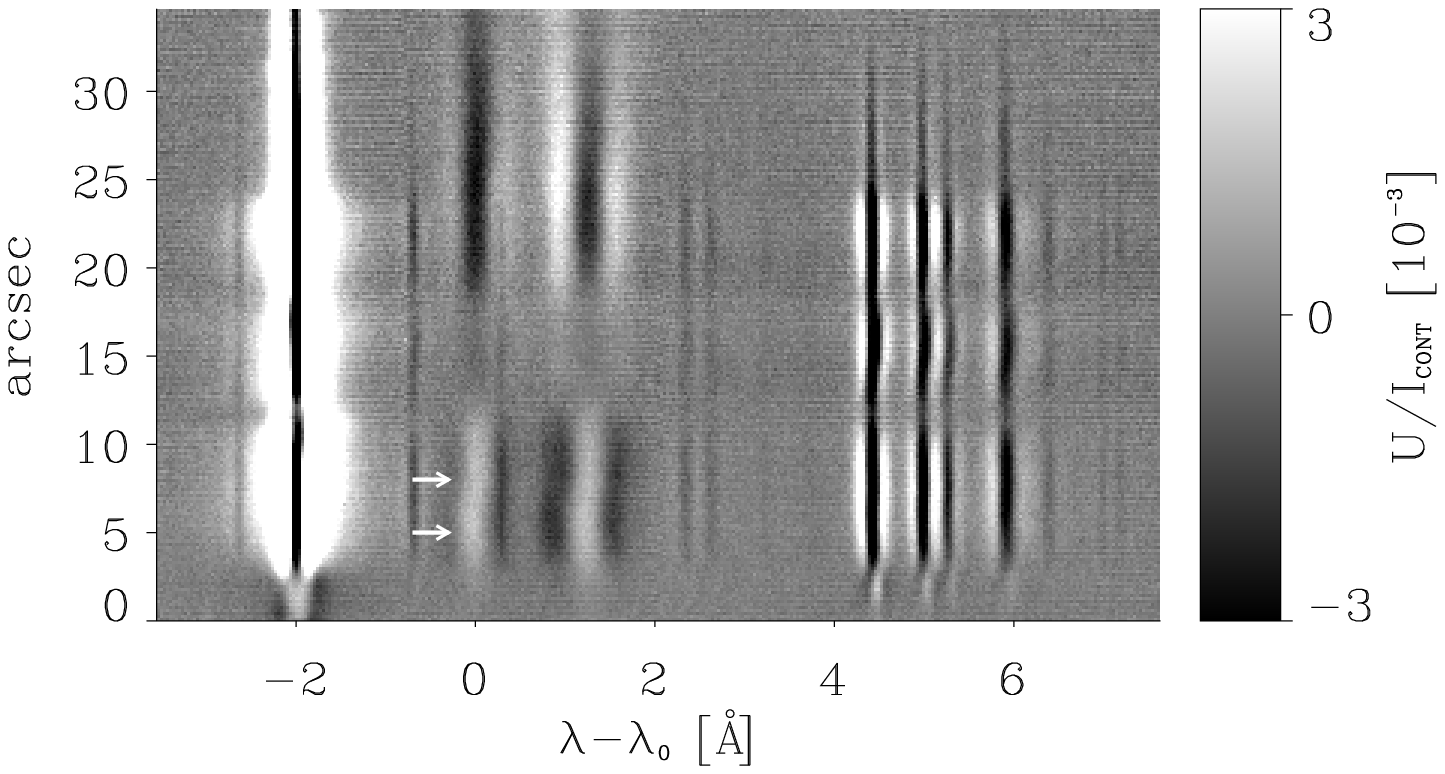}
\includegraphics[width=0.49\linewidth]{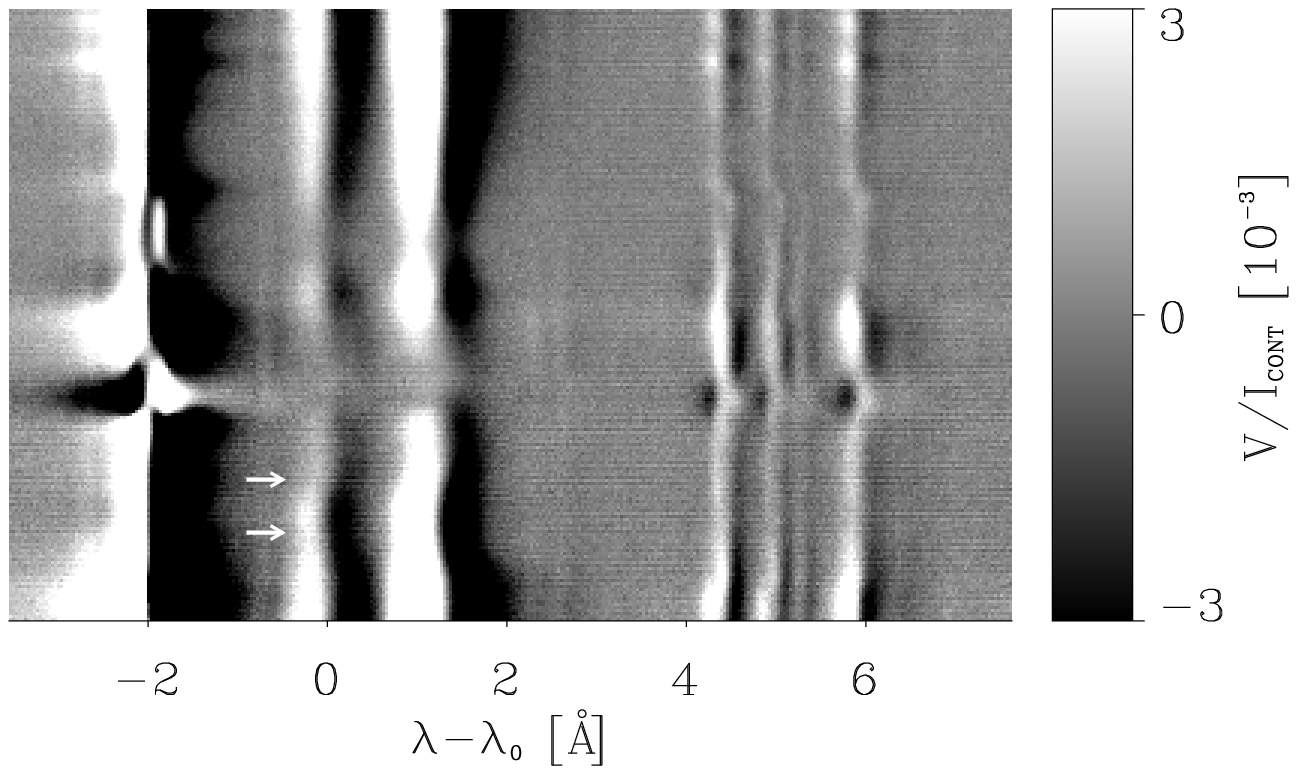} 
\caption{Top left: Slit reconstructed continuum intensity map centered at the NL
where the pores and penumbral-like formations can be identified. The
location of the slit for the time series map is displayed. Top right: \mbox{Si {\sc i}} Stokes V map normalized to the continuum intensity. Bottom: Stokes
profiles observed at the slit position indicated in the top images and averaged over the
time series (comprising 100 scans). The gray scale bar only applies to the polarization 
signals. The Zeeman-like signature of Stokes $Q$ and $U$
are evident in this picture (\mbox{He {\sc i}} lines are centered at 0 and 1.2 \AA\
approximately. The zero in the wavelength scale corresponds to 10829.09 \AA). The arrows indicate the position of the presented Stokes profiles in Fig. \ref{fig:PBinversion1} and Fig. \ref{fig:PBinversion2}.  
}
\label{fig:continuum_with_slit}
\end{figure*}

\section{Observations}

The observations described in this paper were carried out at the German Vacuum
Tower Telescope (VTT, Tenerife, Spain) on the 3rd and the 5th of July, 2005, using
the Tenerife Infrared Polarimeter (TIP-II, \cite{tip2}). TIP-II allows to measure
(almost) simultaneously the full Stokes vector for all the pixels along the
spectrograph slit. The slit (0\farcs5 wide and 35\arcsec\ long) was placed over
the target, a filament lying over the NL of active region NOAA 10781 situated
close to disk center (at coordinates N13-W4 around $\mu=0.97$ on July 3rd and
at N13-W29 or $\mu=0.92$ on July 5th) with the help of context H$\alpha$ and
continuum images. The H$\alpha$ frames showed the filament lying always on top
of bright plage regions visible immediately to either side of the absorption feature.
It is likely that the results presented in this paper particularly apply to
such an ``active'' filament configuration. SOHO/MDI (\cite{scherrer95})
magnetograms and continuum frames have been used to follow the evolution of the
active region as it crossed the visible surface.  The active region was
identified to be in its slow decay mode, encompassing a round symmetric leader
sunspot and a follower, spotless, plage region. At the location of the NL,
sporadic pore-like formations and penumbral-like configurations (with no evident
radially symmetric link to an umbral or pore structure as normal penumbra or partial 
penumbrae do) were seen, especially on the 5th of July data-sets. Pores near the NL with both polarities
are identified and the penumbral-like region seems to correspond to field lines connecting
them. The NL was oriented in the N-E direction and showed a very compact
configuration, with the two polarities always remaining in close contact.  The
leader sunspot became invisible on the 6th of July after decreasing in size
while approaching the West solar limb. 

Two observing strategies were used to map the NL region. Several spatial scans
that covered the whole AR were carried out in the course of the 3 days, and one
time series (with the slit fixed over the filament) was taken on July 5th. This
data-set was averaged in time in order to improve the signal-to-noise (S/N)
ratio of the spectral profiles.  The pixel size along the slit was
0\farcs17 and the scanning step (in the case of the rasters) was 0\farcs3 
per step. The
exposure time per slit position was 8 seconds.  The adaptive optics system
(KAOS, \cite{kaos}) was locked on nearby pores during all the runs,
substantially improving the image quality of the observations carried out during
relatively poor seeing conditions, and providing a final estimated spatial
resolution of $\sim$1\arcsec.  Flat-field and dark current corrections were
performed for all the data-sets and, in order to compensate for the telescope
instrumental polarization, we also carried out the standard polarimetric
calibration (\cite{collados99}; \cite{collados2003}) for this instrument.  

The observed spectral range spanned from 10825 to \mbox{10836 \AA}, with a spectral
sampling of $\sim$11.1 m\AA\,px$^{-1}$. However, a 3px binning in the spectral
domain was applied to all the data to increase the S/N while still preserving
subcritical sampling. The 10830 \AA\ spectral region is a powerful diagnostic
window of the solar atmospheric properties since it contains valuable
information coming simultaneously from the photosphere (carried by the 
\mbox{Si {\sc i}} line at 10827 \AA ) and the chromosphere (encoded in the \mbox{He {\sc i}} 10830
\AA\, triplet). This He multiplet originates between a lower term $2^3S_1$ and
an upper term $2^3P_{2,1,0}$. Thus, it comprised three spectral lines,
namely a ``blue'' component at 10829.09 \AA\, ($J_L=1 \rightarrow J_U=0$) and a
``red'' component at $\sim 10830.3$ \AA\, that results from the remaining two
transitions ($J_L=1 \rightarrow J_U=1,2$), which appear completely blended at
typical solar atmospheric temperatures.
This multiplet is formed in the high chromosphere (\cite{avrett}) with no 
contribution from photospheric levels, and serves as a unique diagnostic tool for
chromospheric magnetic fields. In filament structures, the height of formation of this
multiplet corresponds to the height of the opaque material inside them, which might
well correspond to typical coronal heights. 

\subsection{Predominance of Zeeman-like signatures in AR filament He {\sc i} 10830\,\AA\
Stokes profiles}

The most striking finding encountered during the analysis of the data from this
campaign was the ubiquitous presence of Zeeman-like signatures in the Stokes
$Q$ and $U$ profiles of the \mbox{He {\sc i}} lines in the AR filament
(see Fig. \ref{fig:continuum_with_slit}). Recent
observations of this triplet carried out with the same instrument (Trujillo Bueno et al. 
2002) have shown quiescent filament Stokes profiles dominated by atomic
level polarization (and its modification through the Hanle effect). These 
forward-scattering signatures 
(see Fig. 4 of Trujillo Bueno et al. 2002) correspond to one-lobe profiles that are positive
for the red component of the He triplet and negative for the blue one (in the positive Stokes $Q$
reference system) due, respectively, to selective emission and absorption 
processes induced by the anisotropic illumination of the He atoms. 
However, in the present case, the Stokes $Q$ and $U$ signals of the blue and red components
of the multiplet, exhibit the usual three-lobe profile that is expected from
the Zeeman effect (right panel of Fig. \ref{fig:continuum_with_slit}). 
While some influence of atomic polarization in these profiles cannot be ruled out a priory, 
it is clear that an explanation of their shapes should rely mainly on the Zeeman and 
Paschen-Back effects. This fact, interesting in itself, will be briefly discussed below
(Sect. 3.3).

The continuum frame in the top panel of Figure \ref{fig:continuum_with_slit} was
reconstructed from the various slit scan positions (thus reflecting the real spatial
resolution of the data) of a map centered at the NL. This data-set was taken on the 5th of 
July, 2005. The presence of pores and penumbral-like structures
is evident in this frame and their location corresponds to the NL. Right after this map,
a time series with the slit fixed over the NL
was performed in order to produce high S/N Stokes profiles. 
The bottom four panels of Fig. \ref{fig:continuum_with_slit} correspond to the 
Stokes parameters obtained after averaging over the full time series
(100 scans), which resulted
in a S/N of about 4000. The inversions performed later in this work were 
carried 
out on the spectral profiles extracted from this time-averaged data-set. The location of the NL 
had to be guessed in real-time during the observations. The chosen slit position for the 
time series is represented 
by a vertical black line in the continuum frame of Fig. \ref{fig:continuum_with_slit}. 
Posterior analysis has 
shown that this location was a few arcsec off at one side of the actual NL, and that the 
slit would have been more correctly placed at $x=18$\arcsec\ 
(referred to the abscissa coordinates in the continuum panel) 
instead of at $x=20$\arcsec. Nevertheless, indications of polarity changes along the slit 
in the \mbox{Si {\sc i}} line (evident around $y=13$\arcsec\ in the Stokes $V$ 
map of the same 
figure) show that we were not very far away from it.

Inspection of the profiles at other slit locations, and of similar maps taken on the 3rd 
of July, consistently exhibited Zeeman-dominated linear polarization signals in those regions 
where the He line showed strong absorption features. However, in regions with weak He absorption
(typically near the boundaries of the filament), we often find linear profiles dominated by
scattering polarization signatures, similar to those observed by \cite{trujillo02}. 
This indicates that, in general, the profiles obtained in AR filaments can
have significant contributions from various competing physical processes: 
the atomic level polarization due to
the anisotropic illumination of the He atoms, the modification of these population imbalances  
induced by the presence of a magnetic field inclined with respect to the axis of symmetry 
(Hanle effect), and the Zeeman splitting characteristic of strong magnetic fields. 
A detailed study with these physical ingredients for all the maps observed at this NL is 
beyond the scope of this paper. In the present work, we concentrate on the implications of 
the clear Zeeman-dominated signatures observed almost everywhere in this filament.

One point is worth mentioning after a simple visual inspection of Fig.
\ref{fig:continuum_with_slit}.  A comparison of the signs displayed by the
linear polarization profiles of the \mbox{He {\sc i}} and \mbox{Si {\sc i}} lines shows that
the orientation of the vector magnetic field is different in the filament than
in the underlying photosphere. In the Stokes $Q$ frame, the \mbox{He {\sc i}} 3-lobed
profiles have the same sign all along the slit whereas the \mbox{Si {\sc i}} line (and all the other smaller photospheric features at redder wavelengths)
exhibits a change in sign around $y=23$\arcsec. Conversely, in the Stokes $U$ frame, the
photospheric signals have a 3-lobed sign distribution that remains constant
almost all along the spatial domain while the \mbox{He {\sc i}} lines now show a sign reversal
near the middle of the slit. An analysis based on the formula for the azimuth angle by \cite{auer} gives an estimate of this difference between the azimuths inferred from the \mbox{Si {\sc i}} and \mbox{He {\sc i}} lines. We obtain values of the azimuth around $y = 5$\arcsec\ , which differ in $\sim 50^{\circ}$. This readily shows that the 10830 \AA\ spectral
region has a great potential to diagnose the orientation of the magnetic field
from the photosphere all the way up to the filament. We postpone this study to
a future paper.

\section{Vector magnetic field near the AR neutral line.}

Several analyses with various levels of complexity have been performed on the reduced data. 
The first approach was a simple magnetograph-like analysis based on the
assumption of the weak-field approximation as formulated below.  This
method was applied to all the points in one of the maps obtained
during the campaign. We subsequently performed an analysis of the high S/N spectral profiles
obtained from the averaged time series with 
a Milne-Eddington (ME) code and later with a more sophisticated inversion procedure 
based on Principal Component Analysis (PCA) of a statistically generated database of 
spectral profiles that account for the physics of atomic level polarization and the Hanle effect. 
All these different methods consistently yield transverse field strengths in the filament 
well above 500 G.

\subsection{Magnetograph analysis}

Typical Doppler widths for the red He {\sc i} line are in the range of 200-300 m\AA. The Land\' e
factors are 1.75 and 1.25 for the $J_L=1 \rightarrow J_U=1$ and the $J_L=1 \rightarrow J_U=2$ 
transitions, respectively. If the transitions are weighted with their line strengths,
an average Land\' e factor of $g_\mathrm{eff}=$1.42 for the red component of the Helium multiplet 
is obtained. This Land\' e factor translates the above
Doppler widths to field strengths in the range of 2000-3000 G, 
which is much stronger than the fields we expect for AR filaments. Together with the 
assumption that the observed
signals are due to the Zeeman effect, these large Doppler widths justify the use of the
well-known weak field approximation (see, e.g., \cite{landi92}) as a first approach to 
infer the magnetic field from the data.
In this approximation, the relation between the longitudinal field strength and Stokes
$V$ profile is given by:
\begin{equation}
V_o=f V_m= - f C B \cos\theta {{dI_m}\over{d\lambda}}.
\label{vmag}
\end{equation}
The subscript ``$o$'' stands for the observed profile, while ``$m$'' represents the profile generated 
in the magnetic component that fills a fraction $f$ of the resolution element; 
$B$ is the field strength, $\theta$
the angle between the line-of-sight (LOS) and the magnetic field direction and $C$ the 
constant $4.67 \times 10^{-13} g_\mathrm{eff}\lambda_o^2$ (which forces the wavelength to be
expressed in \AA~and the field strength in G). The observed Stokes $I$ profile is given by:
\begin{equation}
I_o=f I_m +(1-f) I_{nm}
\end{equation}
with the subscript ``${nm}$'' referring to the non-magnetic component.  
The last factor in the right-hand-side of Equation (\ref{vmag}) is the derivative of $I_m$
with respect to the wavelength. However, observations only provide the compound
profile $I_o$, product of the coexistence of magnetic and non-magnetic components in
the same resolution element. The present He {\sc i} analysis benefits from the
fact that the non-magnetic areas of the Sun display a very weak He absorption, so the
derivative of the Stokes $I_{nm}$ profile with respect to $\lambda$ can be neglected 
(we note that, in non-magnetic regions, $dI_{nm}/d\lambda$ is found to peak at 
one order of magnitude smaller values than
$dI_{m}/d\lambda$ as observed in the filament). This nicely eliminates
any dependence of the inferred longitudinal field on the unknown filling 
factor. Following this argument of ignoring the derivative of the $I_{nm}$ profile 
with respect to wavelength, the longitudinal field is directly inferred as:
\begin{equation}
B_{\parallel} =B \cos\theta= - {{1}\over{C}} 
{{V_o(\lambda)}\over{{{dI_o}\over{d\lambda}}(\lambda)}}
\label{blos}
\end{equation}
The $\lambda$ dependence in the last ratio is written explicitly to emphasize that each point
within the profile provides an estimate of the longitudinal field. Thus, the way to
solve Equation (\ref{blos}) is through a least-square fit to all the observed
points within the profile.  

\begin{figure*}
\centering
\includegraphics[width=6cm]{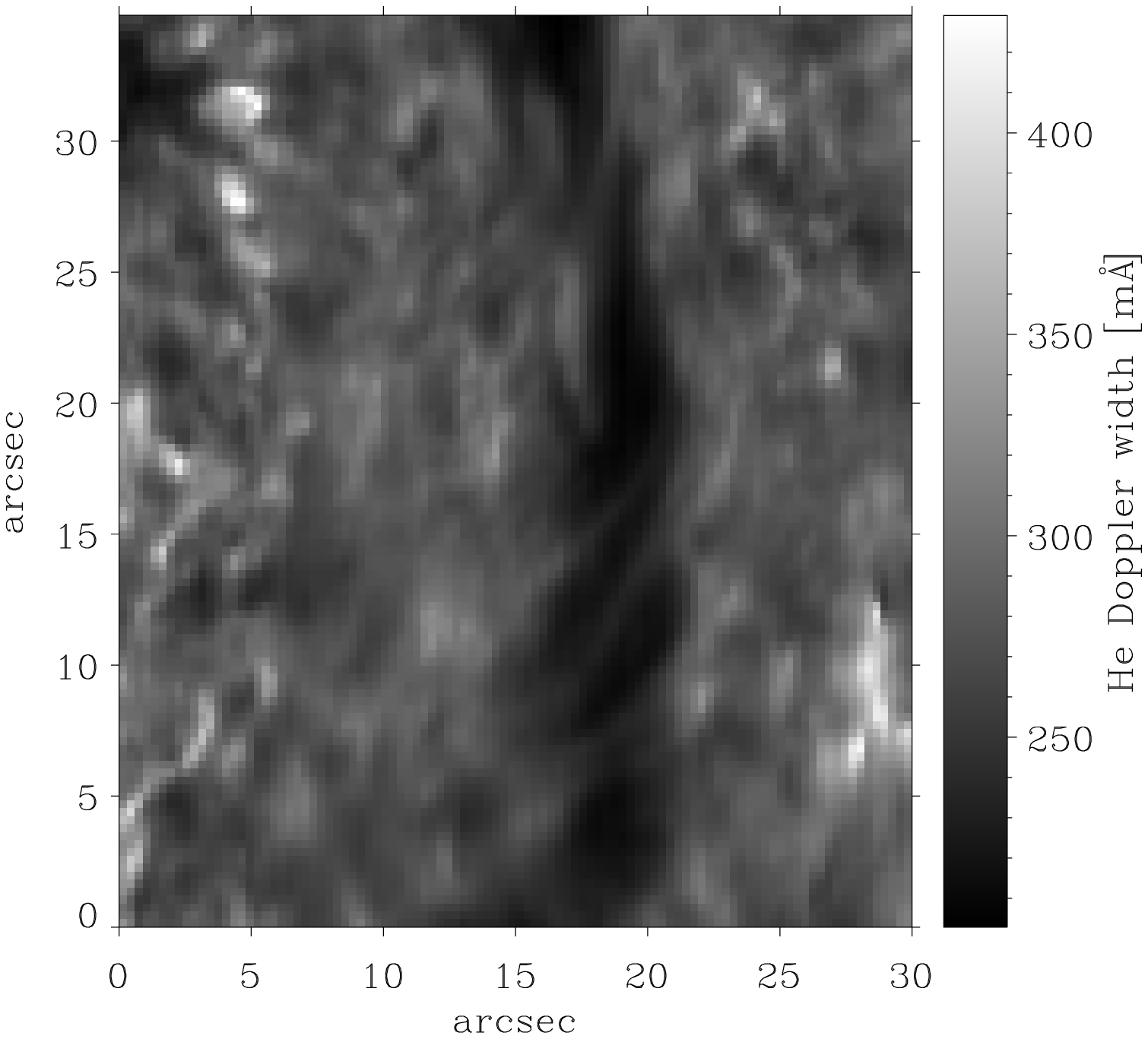} 
\includegraphics[width=6cm]{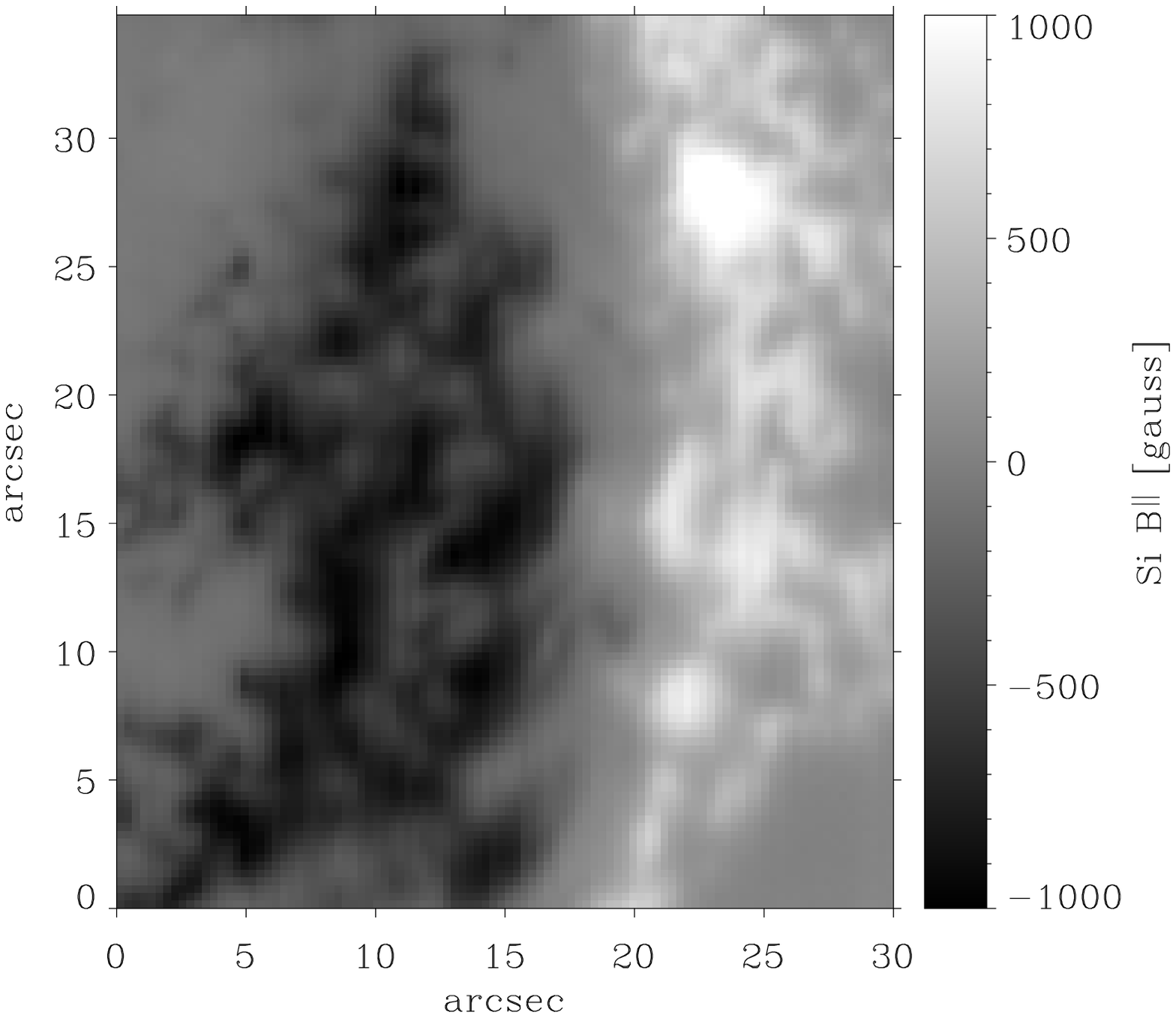} 
\includegraphics[width=6cm]{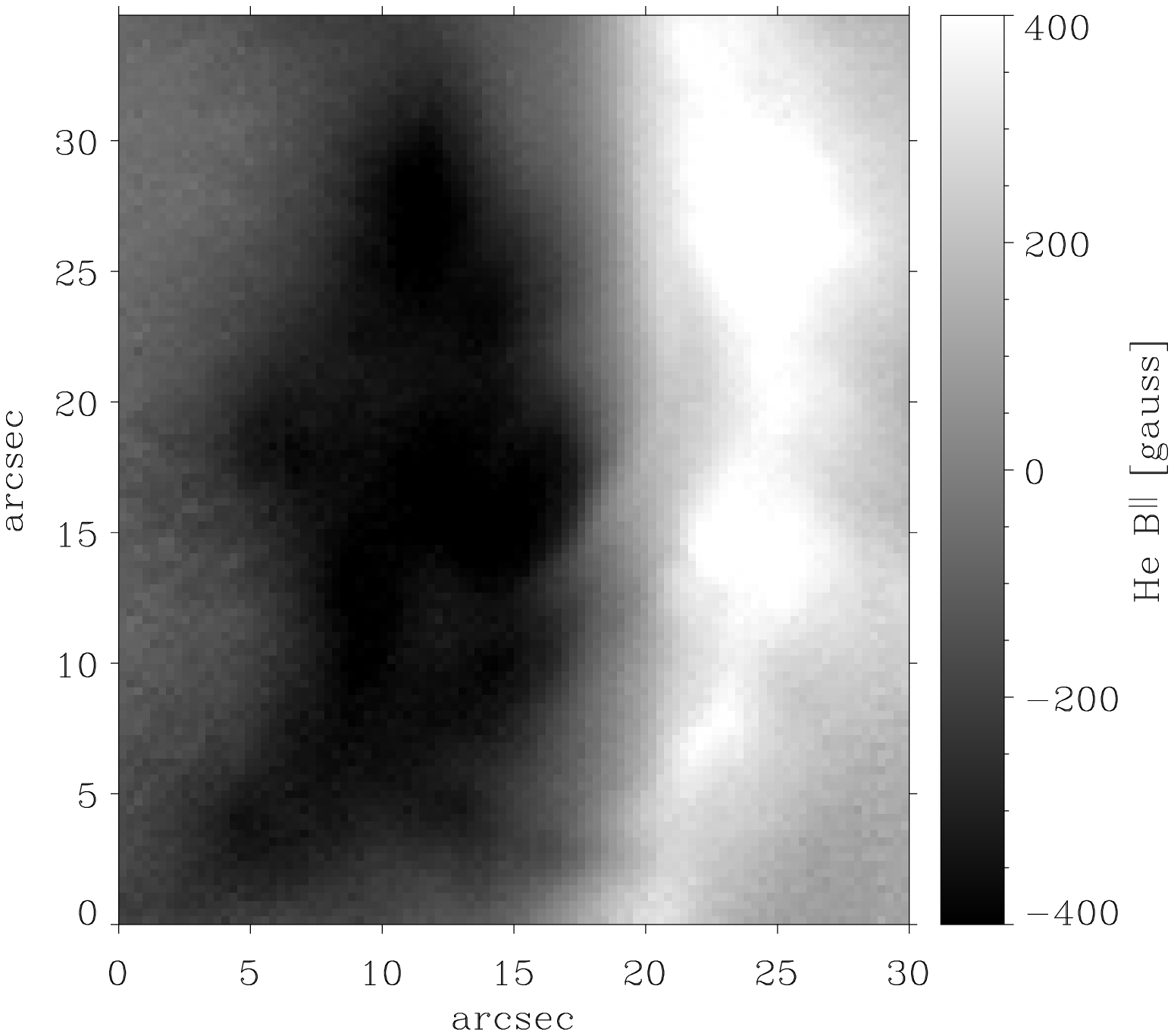} \\
\includegraphics[width=6cm]{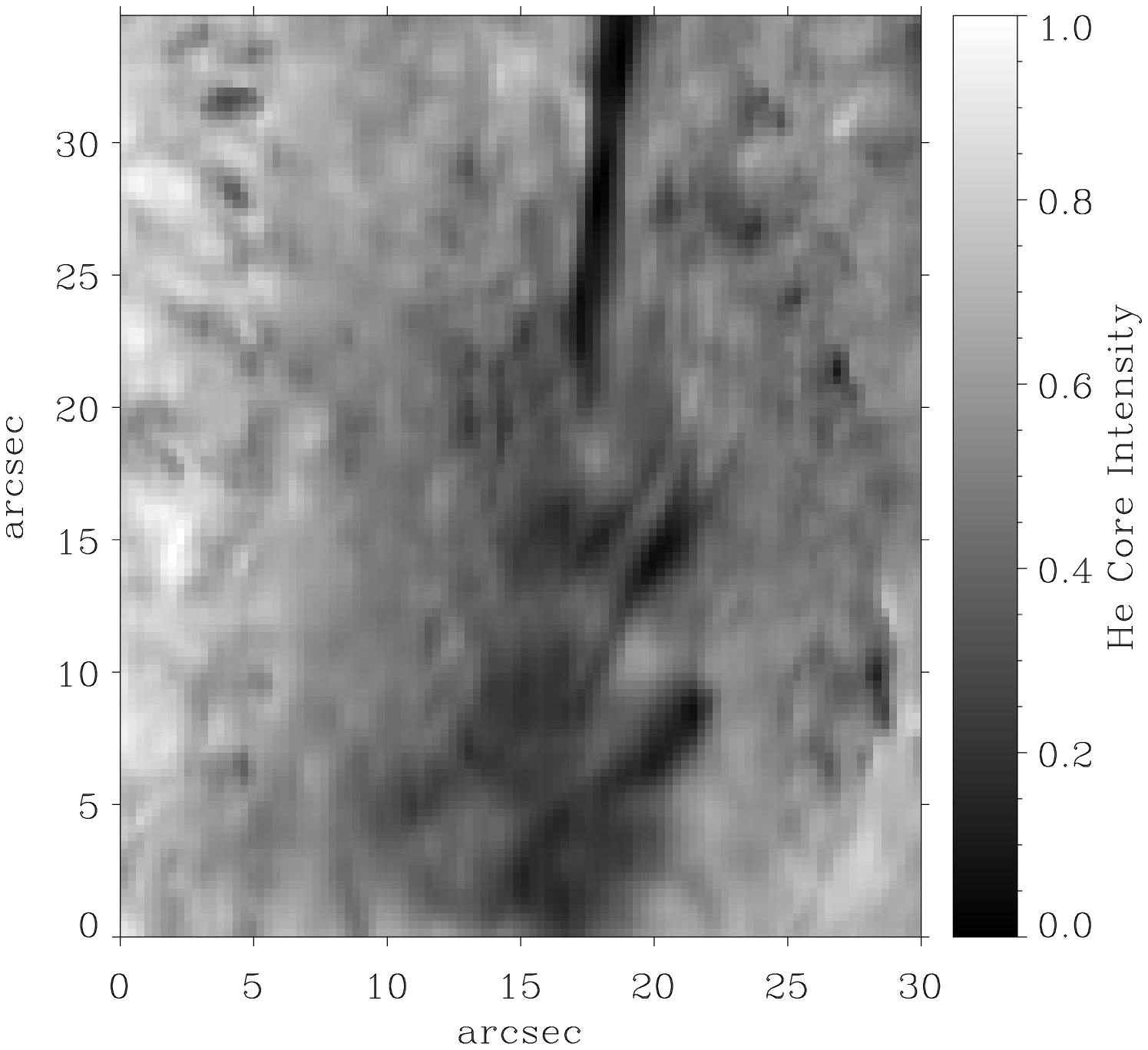} 
\includegraphics[width=6cm]{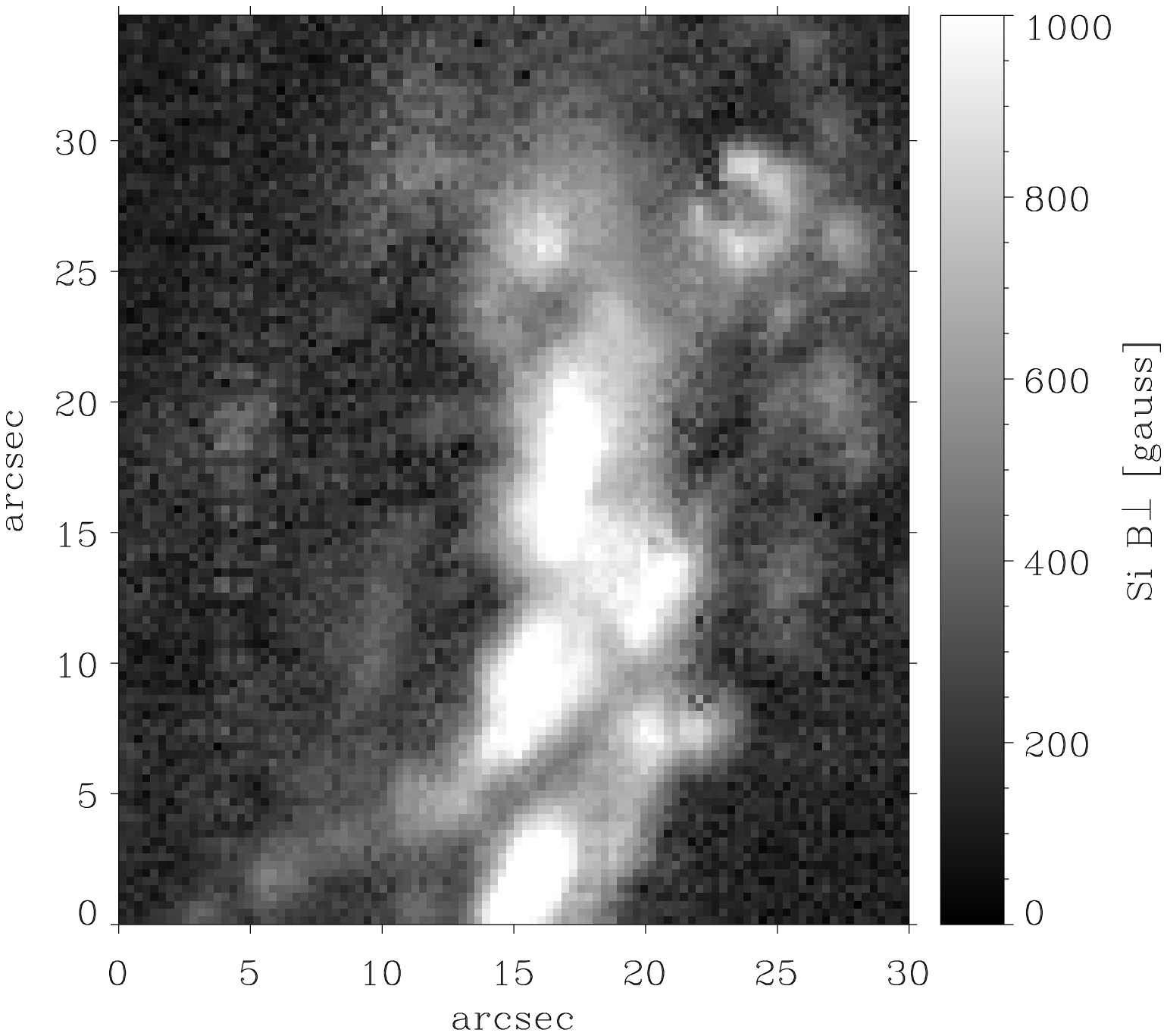} 
\includegraphics[width=6cm]{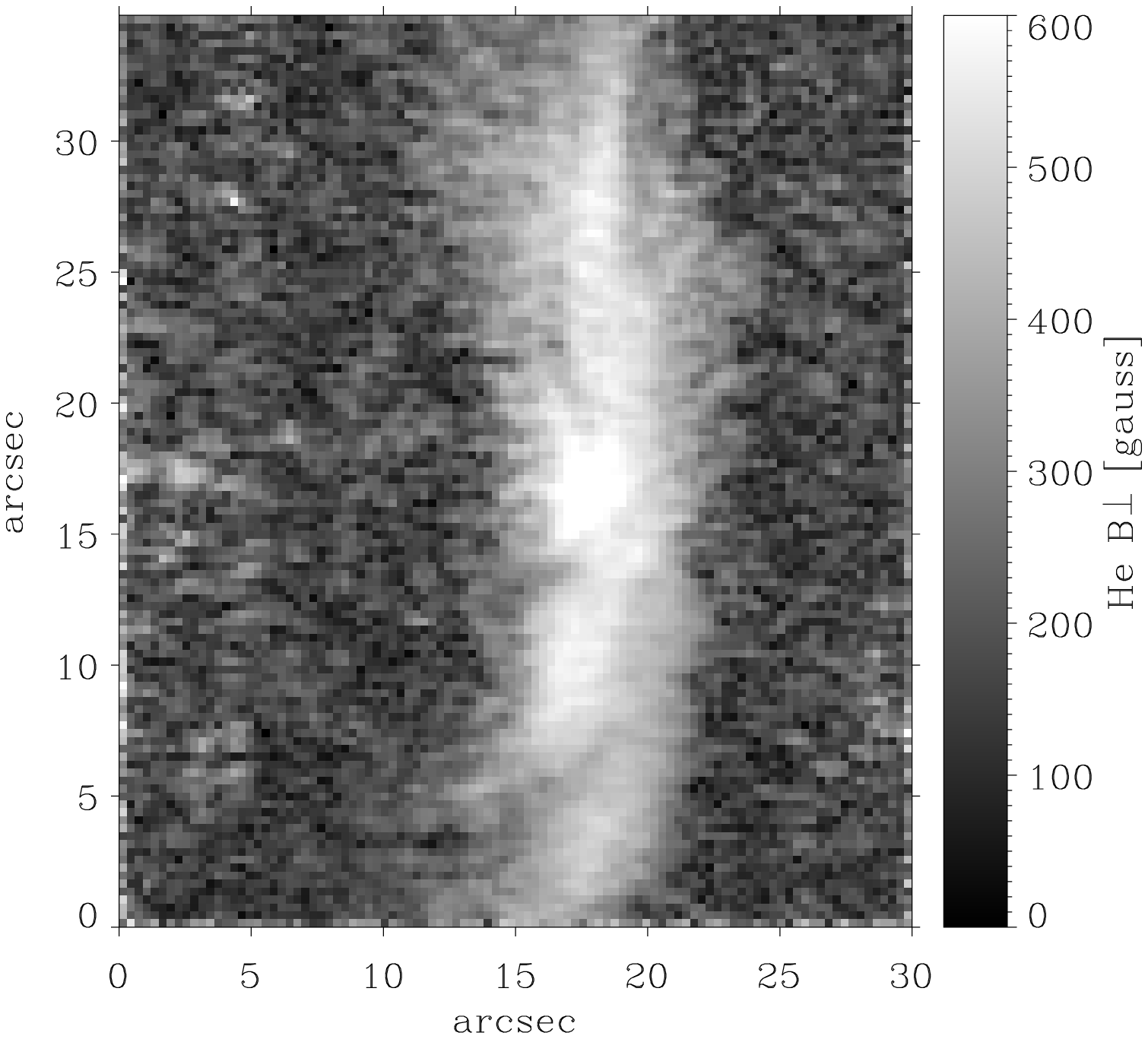}
\caption{From left to right and top to bottom: He {\sc i} Doppler width, \mbox{Si {\sc i}} LOS magnetic
field, He {\sc i} LOS magnetic field, He {\sc i} core intensity, \mbox{Si {\sc i}} transverse magnetic
field and He {\sc i} transverse magnetic field. The corresponding continuum intensity
frame can be seen in the top panel of Fig. \ref{fig:continuum_with_slit}.
}
\label{fig:magneto}
\end{figure*}
For transverse fields, a similar equation applies including 
the quadratic dependence of the linear polarization signals on
the transverse component of the field 
and the second derivative of the Stokes $I$ profile with respect to $\lambda$.
A similar argument gives:
\begin{equation}
B^2_{\perp} =B^2 \sin^2\theta=
{{4}\over{C^2}}{\sqrt{Q^2_o(\lambda_o)+U^2_o(\lambda_o)}
\over{{{d^2I_o}\over{d\lambda^2}}(\lambda_o)}}.
\label{btra}
\end{equation}
Like in the equation for the longitudinal field, the filling factor (of a
non-magnetic component) is absent.  On the other hand, and in contrast to Eq.
(\ref{blos}), the right hand side of Eq. (\ref{btra}) must be computed at the
central wavelength, $\lambda_o$, as this equation can only be formally derived
there (\cite{landi92}). Having only one wavelength point to compute the
transverse fields prevents the use of a least square approach, unlike in the
case of the longitudinal field equation. This, in turn, complicates the
determination of the transverse component because the Stokes $Q$ and $U$ signals are
often found to be at, or close to, the noise level. The quadratic average of
the linear polarization signals in \mbox{Eq. (\ref{btra})} hinders any possible
noise reduction and generates a veil of noise-induced transverse fields that
proves problematic in the use of this equation in many cases.
This is particularly true for the He {\sc i} line, whose second
derivative of the (broad) Stokes $I_o$ profile is very small. This
derivative in the denominator in \mbox{Equation (\ref{btra})}
results in values of the noise-induced transverse fields as large as
300-400 G, only slightly smaller than the actual values
measured in the filament region. Note that this 300-400 G noise level
corresponds to a S/N of 1000 in the map in \mbox{Fig. \ref{fig:magneto}}, whereas in
\mbox{Fig. \ref{fig:continuum_with_slit}} the S/N is four times larger. 

Thus, and in order to obtain high S/N maps of the chromospheric 
transverse fields, we averaged the unsigned $Q$ and $U$ signals over
three wavelengths, the central wavelength and two wavelengths at 278 m\AA\ 
on each side of line center, corresponding approximately to the locations of the peak 
signals in the linear polarization profiles, so 
\begin{equation}
\sqrt{Q^2_o(\lambda_o)+U^2_o(\lambda_o)}\approx\sqrt{{\bar Q}^2_o+{\bar U}^2_o},
\label{btra2}
\end{equation}
where ${\bar Q_o}$ and $\bar{U_o}$ are the averaged linear polarization signals over
these three wavelength points. We emphasize that this approach reduces
the noise level considerably 
while at the same time preserves the distribution of transverse fields
already present when directly applying Eq. (\ref{btra}). The results
obtained from this approach have been validated by the ME inversions presented
in the next section.

A similar approach was followed with the \mbox{Si {\sc i}} line to obtain the
photospheric magnetic field. The Land\' e factor for this line is 1.5. 
The spectral profiles arising from this transition are broad enough to partly justify 
the application of the weak-field approximation. However, like for any other photospheric line,
the vector-magnetograph data obtained with this method have to be interpreted with caution. 
In particular, the results for this line are affected by filling factor effects as is the 
case for commonly used photospheric 
magnetograph data. For the \mbox{Si {\sc i}} line, the exact formulation of Eq. (\ref{btra}) was 
used. The intrinsically larger photospheric magnetic field strengths together with the 
fact that the \mbox{Si {\sc i}} Stokes $I$ profile has a sizeable second derivative, makes 
the determination of $B_{\perp}$ less affected by noise when using only the central wavelength 
point of the linear polarization profiles for its determination. 

Lastly, the Stokes $I$ profiles 
of both \mbox{He {\sc i}} and \mbox{Si {\sc i}}  were fitted with 
gaussian functions from which the line center, line width and strength were inferred. 
In the case of the \mbox{He {\sc i}}, this approach proved very useful to identify the location of the
filament above the NL. 

The results from this approach are presented in Figure \ref{fig:magneto} for
the map observed on the 5th of July. The He {\sc i} Doppler width and line core
frames show that the line becomes deeper and narrower inside the filament. This
could be an indication of the presence of a denser and cooler plasma than in
its immediate plage surroundings.  These frames also reveal that the filament
has a highly twisted topology, with filamentary threads running at almost 45
degrees from the direction defined by the NL. This becomes evident when
compared to the Si and He LOS magnetograms.  The twist in the filament is more
obvious in the bottom half of the frames than in the top part, where a more
diffuse linear topology is observed. These twisted signatures could be clearly
seen in the observations from the 5th of July, while they were hardly visible
in the maps obtained two days earlier. The \mbox{Si {\sc i}} LOS magnetogram
shows that the separation between the two polarities is less than 5\arcsec at
photospheric levels when the magnetogram is scaled at $\pm$1000 G. A tighter
scaling with a smaller threshold would provide a much narrower NL channel.  In
the \mbox{He {\sc i}} LOS magnetograms, with a scaling of $\pm$400 G, the
channel is practically absent, showing how intense the plage surrounding this
AR filament was. A thread-like structure in the longitudinal component running
parallel to the direction of the twist (i.e. at 45 degrees from the NL) in the
mid range of the frame is observed. Note that the AR is 22 degrees off disk
center so horizontal threads in the filament can give rise to sizeable Stokes
$V$ signals.  The intrinsic LOS magnetic field strength at the NL (we stress
that no filling factor effect is included in these estimates) is typically 100
G for each of the two polarities (in agreement with older measurements of AR
filaments that did not include transverse field measurements).  The neighboring
plage displays an intrinsic longitudinal field strength of 200-400 G. If we
assume an intrinsic photospheric field strength in the plage of \mbox{1400 G} (see
\cite{martinezpillet97}), this would indicate a filling factor of about 20\% in
this layer. 

In both the photospheric and chromospheric magnetograms, the region where the
longitudinal field becomes weaker corresponds to the region with very strong
transverse fields. The somewhat wider photospheric NL channel corresponds to a
wider photospheric area occupied with strong transverse fields, while the
transverse fields in the chromosphere fill a narrower region. These transverse
fields seem to follow the structures observed in the He line core frame: a more
linear topology in the top part and a twisted formation in the bottom half.
What becomes readily surprising is the magnitude of the transverse fields
observed in the chromosphere (and to the extent that the weak field
approximation is valid, in the photosphere as well). The photosphere shows
fields in the range of up to kG in the region between abscissae $x=15$\arcsec\
to $x=20$\arcsec\ in the frames of Fig. \ref{fig:magneto}.  This area shows
penumbral-like structures and pores in the white light image.  It is clear
that these orphan penumbral-like regions are magnetically linked to the main body of
the AR filament and correspond to horizontal field lines at the photospheric
surface. The transverse fields derived from the \mbox{Si {\sc i}} line
undoubtedly show the twist configuration similar to that observed in the
\mbox{He {\sc i}} lines.  In the chromosphere, the transverse fields are also
strong. The present analysis yields transverse fields with strengths in the
range of 500-600 G, strongly concentrated in the narrow NL channel. Note that
this region, at best, a few arcsec wide, is below the low resolutions
of old observations. These large transverse magnetic fields are well above
most, if not all, of past field strength measurements in AR filaments. 

The various maps for the observations on the 3rd of July support the hypothesis
that an AR filament corresponds to a narrow channel characterized with strong
($\sim 500$ G) transverse fields.  There is a clear spatial correlation between
strong absorption signatures in \mbox{He {\sc i}} (measured by the line core
intensities), and narrow (or ``abutted'') plage regions with strong transverse
fields in the chromosphere.  A complete description of all these maps is
postponed to a future paper. But it is interesting to point out the
following main differences with the map of Fig. \ref{fig:magneto}: the absence
of evident twisted threads, the absence of penumbral-like regions in
white-light maps, weaker and more diffuse transverse fields at the photosphere
(not reaching kG levels) and the presence of weaker \mbox{He {\sc i}}
absorption signatures in regions outside the intense NL channel that display
atomic polarization signals.

\begin{figure*}
\centering
\includegraphics[width=14cm]{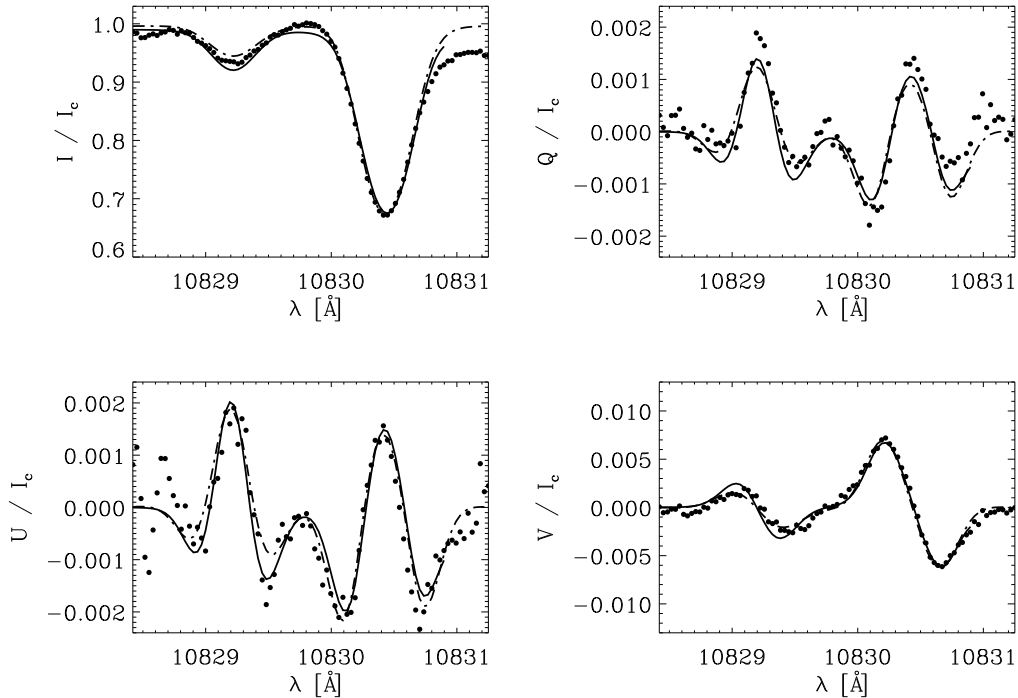}
\caption{Stokes profiles of the \mbox{He {\sc i}} 10830 \AA\ triplet at a height
of 8 arcsec in Fig. \ref{fig:continuum_with_slit} 
in the mid portion of the twisted region of the
filament. The dots represent the observed profiles obtained after averaging
the time series in order to achieve a larger S/N. The solid line corresponds
to the best fit achieved with a Milne-Eddington inversion code that takes
into account the IPB effect. The magnetic field obtained from
this particular fit is $B = 626\pm 10$ G,
$\theta = 80.2^{\circ}\pm 0.1^{\circ}$ and
$\chi =174.9^{\circ}\pm 0.1^{\circ}$ (the origin of azimuths being the local
solar radial direction).
The dot-dashed line shows the best fit obtained from an inversion in a slab
model of constant properties that accounts for the joint action of scattering 
polarization 
and the Hanle Zeeman effects.
This fit provided for the magnetic parameters 
$B = 667\pm 16$ G, $\theta = 81.1^{\circ}\pm 0.3^{\circ}$ and $\chi = 174.8^{\circ}\pm 0.5^{\circ}$.
}
\label{fig:PBinversion1}
\end{figure*}

\begin{figure*}
\centering
\includegraphics[width=14cm]{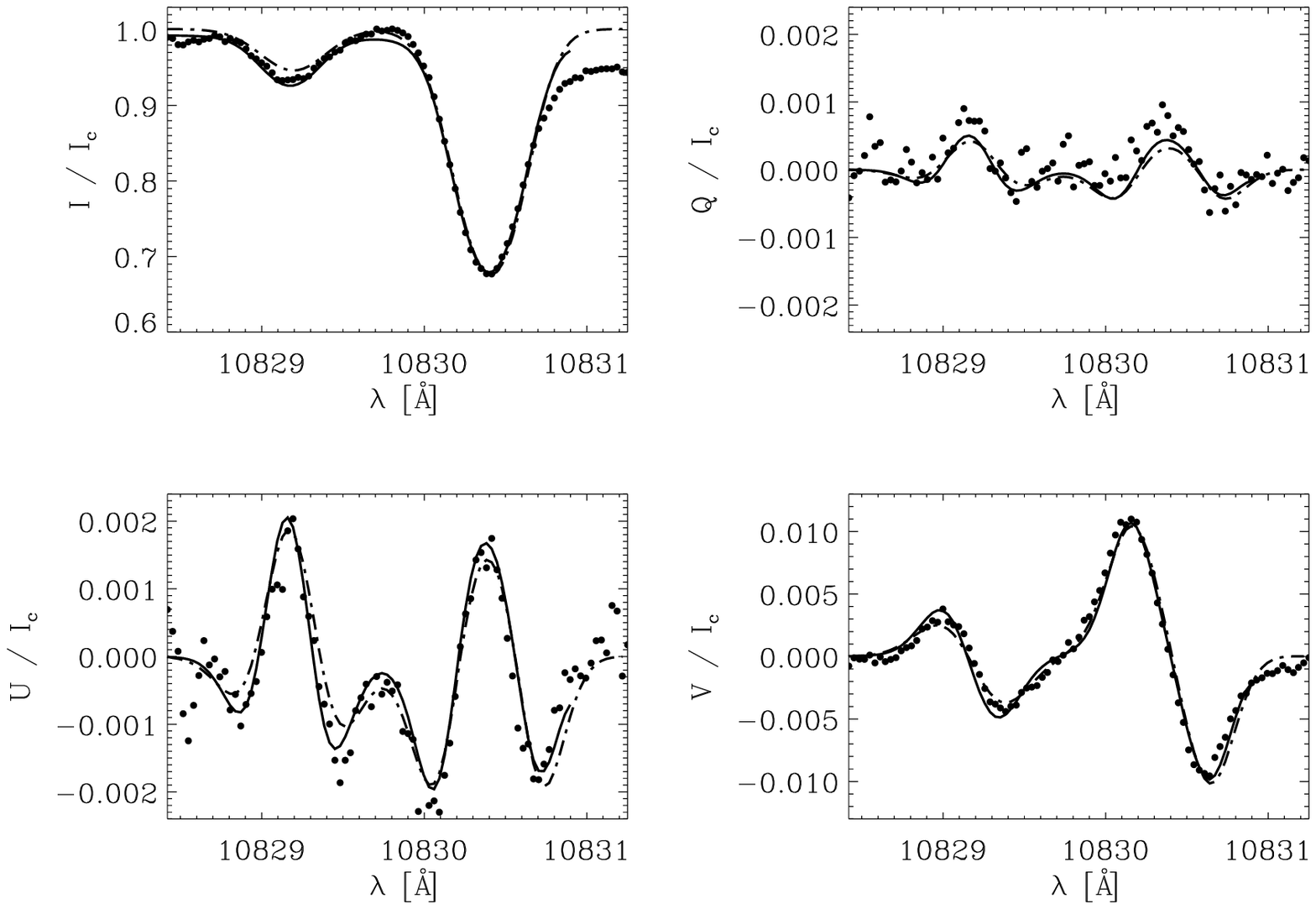}
\caption{
Same as Fig. \ref{fig:PBinversion1} but for a profile near the bottom portion of the
twisted region of the filament (at a height of 5 arcsec). 
The fitted parameters for the ME case are
$B = 627\pm 10$ G,
$\theta = 73.7^{\circ}\pm 0.3^{\circ}$ and
$\chi =5.7^{\circ}\pm 0.1^{\circ}$. The PCA-based algorithm including Hanle effect provided:
\mbox{$B = 664 \pm 15$ G}, $\theta = 74.9^{\circ}\pm 0.6^{\circ}$ and $\chi = 5.1^{\circ}\pm 0.8^{\circ}$.
}
\label{fig:PBinversion2}
\end{figure*}

\subsection{Milne-Eddington inversions}
The interest of the magnetograph analysis is that it is easy
computed over a complete map with a very low computational effort. However, in 
order to validate the weak-field approximation inferences of the magnetic
field strength in the filament, we carried out a Milne-Eddington (ME) inversion
of the \mbox{He {\sc i}} Stokes profiles of the averaged time series. These
high S/N data correspond to the location of the vertical line at $x=20$\arcsec\
in the map of Fig. \ref{fig:continuum_with_slit}.  The inversion code we used
(MELANIE; \cite{MELANIE}) computes the Zeeman-induced Stokes spectra---in the
Incomplete Paschen-Back (IPB) effect regime---that emerge from a model
atmosphere described by the Milne-Eddington approximation. This assumes a
semi-infinite constant-property atmosphere whose source function varies
linearly with  optical depth, $\tau$.  This approach does not account for the
atomic-level polarization induced by the anisotropic radiation pumping of the
He atoms from the underlying photospheric continuum.

For magnetic fields between 400 and 1500 G the Zeeman splitting of the upper
J-levels of the \mbox{He {\sc i}} triplet is comparable to their energy separation. Thus, it
is crucial to compute the energy levels in the IPB regime to avoid an
under-estimation of the magnetic field strength when carrying out the inversion
to interpret observations (\cite{sn-pb}; \cite{sasso}).

The Milne-Eddington model atmosphere uses a set of eleven free parameters that
the inversion code modifies in an iterative manner in order to obtain the best fits to 
the observed Stokes profiles. These free parameters are: the magnetic field strength ($B$), 
its inclination ($\theta$) and azimuth ($\chi$) angles in the
reference system of the observer (the azimuth is measured
with respect to the local solar radial direction), the line strength ($\eta_0$), 
the Doppler width ($\Delta\lambda_D$), a damping parameter, the LOS velocity, the source 
function at $\tau=0$ and its gradient, a macroturbulence factor and a stray light 
fraction $f$.
An initial guess model must be provided to the code. In order to prevent the inversion scheme
from getting locked into local minima, we carried out the inversions using several different 
initializations, in which $B$, $\theta$ and $\chi$ were taken from 
the above-described magnetograph analysis and the remaining parameters were obtained 
from random perturbations of reasonable pre-set values.
All parameters were free except for the stray light fraction, which was set to zero 
(i.e., assuming a filling factor of 1; we refer to the explanation in section 3.1).
We stress that, even if the data is contaminated with stray-light from the surroundings,
using a filling factor of unity allows a direct comparison with the results 
given in the previous section. 

Figures \ref{fig:PBinversion1} and \ref{fig:PBinversion2} 
show the results of the ME inversion of the full Stokes 
vector for two positions along the filament (only two
fits are shown but the complete set of profiles from 
Fig.~\ref{fig:continuum_with_slit}
were inverted). The excellent performance of the ME inversion
tells us that there are barely any scattering polarization signatures in these
profiles, and that the formation physics of the multiplet is adequately described
in the IPB regime. As expected, the magnetic field strengths inferred from all the 
inversions are of the order of 600--700\,G with very high 
inclinations ($\sim 80^{\circ}$)
with respect to the line-of-sight, thus confirming the existence of 
strong transverse
fields in this AR filament. The exact values obtained for 
Fig.~\ref{fig:PBinversion1}
are $B_{\parallel}=109$\,G and $B_{\perp}=616$\,G while for 
Fig.~\ref{fig:PBinversion2} we
obtain $B_{\parallel}=176$\,G and $B_{\perp}=602$\,G. 
In the solar reference frame, the magnetic field
vector turns out to be close to horizontal, 59$^\circ$ and 99$^\circ$ of
inclination with respect to the local vertical, respectively.

The magnetic field strengths derived from the weak field 
regime analysis are systematically under-estimated by \mbox{$\sim 100 - 150$ G}, but 
the retrieved inclinations and azimuths are in very good agreement with those 
obtained from the ME-IPB inversions. This means that the weak field approximation 
yields a reliable magnetic field topology and a low-end value for the field strength.

\subsection{PCA-based atomic polarization inversion}

The observations described above were also independently inverted
using the Principal Component Analysis (PCA) approach described by 
L{\'o}pez Ariste \& Casini (2002). In fact, pattern recognition techniques are particularly
well suited to attack ill-defined inversion problems characterized by
computationally intensive forward problems. This is exactly the case 
for spectropolarimetric inversions in prominences and filaments, 
where the Stokes profiles are often formed by the scattering 
of resonant radiation. The computation of the emergent polarization in
such case requires the preliminary solution of the non-LTE problem of 
atomic-level excitation by anisotropic illumination of the plasma 
from the underlying photosphere. The presence of a magnetic field 
further modifies the ensuing atomic polarization through the Hanle 
effect (see, e.g. \cite{LL04}, for a review of atomic polarization effects). 
This problem, which is totally by-passed in the ME approach 
to spectropolarimetric inversion, completely dominates the numerical 
computation of line polarization in radiation scattering. For this
reason, pattern recognition techniques provide a very attractive 
strategy to Stokes inversion for radiation scattering, since the 
numerically intensive forward problem is solved once and for all for 
a comprehensive set of illumination, thermodynamic, and magnetic 
conditions in the plasma for the problem at hand. The goal of these 
techniques is thus to build universal databases of profiles that can 
be searched for the solution to any given Stokes inversion problem. 
Principal Component Analysis additionally provides a way of 
compressing database information, by reducing it to a few principal 
component profiles that contain all the fundamental physics of the 
formation of the emergent Stokes profiles (\cite{LC02,CBL05}).

For the inversion of the observations illustrated in this paper, we created a
database of 250,000 profiles, spanning all possible orientations of the
magnetic field, with strengths between 0 and 2000\,G. The illumination
conditions were set by the radiation temperature and center-to-limb variation
(CLV) profile of the photospheric radiation at 1$\mu$m (\cite{cox}), and by
assuming a range of scattering heights between 0 and 0.06\,$R_\odot$ (42 Mm).
The LOS inclination in the database used for the inversion of the July 5, 2005,
observations spanned between 20$^\circ$ and 30$^\circ$. The thermal Doppler
width and micro-turbulent velocity, responsible for the overall profile
broadening, were accounted for by introducing an equivalent temperature
spanning between $10^4$ and $2\times 10^4$\,K.  Finally, the database Stokes
profiles were calculated by integrating the polarized radiation emerging from a
homogeneous slab with optical depth at line center varying between 0.2 and 1.4.
Very soon it was realized that such conventional scattering scenario could not
fit satisfactorily the observations, unless some additional depolarizing
mechanism could be accounted for to explain the surprisingly low level of
atomic polarization revealed by the Zeeman-like shape of the Stokes $Q$ and $U$
profiles. For this reason, in the creation of the inversion database we
introduced an ad-hoc weight factor for the anisotropy of the photosphere
radiation, ranging between 0 and 1. Extensive inversion tests that were run
over the entire filament map consistently gave anisotropy weight factors
significantly smaller than unity, with a predominance of values around 0.2,
thus confirming the presence of some unidentified depolarizing mechanism in the
formation of the observed profiles. The possible physical origin of such
depolarization has been extensively discussed by \cite{CML09}, but see also
Trujillo Bueno \& Asensio Ramos (2007). 

The results of the PCA inversion are overplotted to the ME inversions as dot-dashed lines in Figs. \ref{fig:PBinversion1} and \ref{fig:PBinversion2}. The PCA inversion provides a magnetic field of $B=667$\,G in the first case and \mbox{$B = 664$\,G} in the latter. The inclination angles are almost the same as the ones obtained by the ME inversions, \mbox{$\sim 59^{\circ}$} and $\sim 99^{\circ}$ with respect to the local vertical, respectively. The inferred magnetic field azimuths are also in very good agreement.

\section{Conclusions}

The He {\sc i} 10830 \AA~lines observed in a filament at the NL of Active Region
NOAA 10781 exhibit linear polarization profiles dominated by the
Zeeman effect. Three different independent analyses (magnetograph approximation,
ME inversion and PCA inversion including atomic polarization) of the four Stokes 
profiles consistently support inferred field strengths in the range 600-700 G at the formation 
height of the Helium triplet. These fields are 3 to 7 times larger than those measured heretofore 
in AR filaments. The field strengths found at the NL are largely horizontal with 500-600 G
transverse fields. The longitudinal component is typically measured to be in the range 
of 100-200 G, in agreement with past measurements. It must be stressed that these previous observations
of AR filaments did not include transverse field estimates. 
Thus, it is clear that the inclusion of the Stokes $Q$ and $U$ profiles in our analysis is what 
has allowed the detection of such strong magnetic fields in AR filaments. While
the role played by the linear polarization signals was evident already from the magnetograph 
analysis, a further test was made with the ME inversion to prove this point.
Using as input the observed $I$ and $V$ profiles but replacing the Stokes $Q$ and $U$ profiles
with noisy data, the ME inversions resulted in fields strengths in the range of 100-200 G. 
We thus propose that the lack of full Stokes
polarimetry was the main reason why past measurements did not find the high field
strengths reported in this work. Note also that the spatial extent that displays such strong fields
is not more than a few arcsec wide, which would be hardly visible in low resolution
measurements. It is also clear that it becomes crucial to search for the
signatures of these large field strengths in different spectral windows such as H$\alpha$ and
the Ca {\sc ii} triplet. 

A similar trend to infer higher field strengths (albeit in a lower range) is 
presented in modern measurements of quiescent filaments that is also ascribed 
to the use of full Stokes polarimetry (Casini et al. 2003). In the case 
of quiescent filaments, observations are dominated by atomic polarization and its 
modification through the Hanle effect. In the study presented
here, however, 
the linear polarization profiles are dominated by the Zeeman effect as long as the profiles
are coming from the regions of strong Helium absorption. The reason why atomic polarization
signatures are almost absent from these profiles is not yet well understood. On the one hand,
the observed high field strengths are approaching the values where the Zeeman effect dominates
over atomic level polarization (\cite{tb07}), but even at this high field strengths, 
one would have expected a clearer atomic polarization signature. On the other hand, some mechanism
to reduce the anisotropy of the radiation field could be present. For example,
Trujillo Bueno \& Asensio Ramos (2007) suggested that, if the
radiation comes from a high opacity region, the isotropy of the radiation field will
be such that a much reduced atomic polarization would be induced. A more recent proposal by \cite{CML09}
suggests that the presence of a randomly oriented field entangled with the main 
filament field, and of a similar magnitude, could also explain the absence of atomic polarization
signatures in these profiles. It is interesting to note that the ad-hoc anisotropy
weight factor introduced in the PCA inversions was found to be in the range 0.1-0.5
over the filament region, with a predominant value of 0.2. Although the various atomic processes
generating polarization signals cannot be cleanly separated, this persisting value
of 0.2 is evidence of their presence in our profiles
(note that a weight factor of 1 corresponds to an anisotropic illumination described by the
standard CLV of the photospheric radiation field). We also
stress that the reason for the absence of atomic level polarization signatures is not simply due to
what is commonly referred to as a Van-Vleck configuration of the vector field. For example,
whereas the profiles inverted in Fig. \ref{fig:PBinversion1} give an inclination 
(with respect to the local vertical) close to 
the Van-Vleck angle (59$^\circ$; Van-Vleck angle corresponds to $\sim$ 55$^\circ$), those in 
Fig. \ref{fig:PBinversion2} yield an inclination very far from it (99$^\circ$).

What is the origin of these strong transverse fields? This question relates
directly to the problem of filament formation and mass loading. Two basic
scenarios are commonly used to explain how these structures are formed:
photospheric (shearing) motions and flux emergence (see the recent review by
\cite{lites08}).  The first one uses photospheric plasma flows that move,
tangle and reconnect the  field lines of an already emerged active region to
form the filament directly in the corona. These processes include in one way or
another some form of magnetic cancellation and reconnection that provides a
source for mass upload of the filament. In this scenario, the presence of such
strong magnetic fields must be related to the existence of a dense plage
configuration at the NL and the low gradients inferred by
\cite{aulanier&demoulin03} for their model of AR filaments.  Our observations
pose the question of how filament field strengths in the range of 600-700 G may
be generated in this scenario from a surrounding ``abutted'' plage that has a
longitudinal field of no more than 400 G at the height of formation of the
Helium lines. The emergence of a flux rope from below the photosphere scenario has recently received strong
support from  the observations of \cite{okamoto08}. If the flux ropes are formed below the surface,
the answer to the observed field strengths could be related to the balance between 
buoyancy forces and gravity acting over the flux system. Although this
balance is not yet fully understood, it is clear that the stronger the fields the easier
is for the flux system to emerge into the Corona and carry a significant amount of trapped
photospheric mass (\cite{Archontis04}).
Note that the photospheric transverse fields observed at the NL 
are also very strong (including pore-like structures) and could represent the bottom part of 
the flux rope system once emerged into the atmosphere. 

It remains to be studied whether the observed strong transverse field strengths
presented in this paper are common to all AR filaments or only to those
surrounded by exceptionally dense plages. An extension of the present study to
other ARs with different degrees of activity is mandatory.

\begin{acknowledgements}
This work has been partly funded through project ESP2006-13030-C06-01.  The
National Center for Atmospheric Research (NCAR) is sponsored by the National
Science Foundation.  Help received by C. Kuckein during his stay at HAO/NCAR is
gratefully acknowledged. R. Manso Sainz has been partially supported by Spanish
Ministerio de Ciencia e Innovaci{\' o}n through project AYA2007-63881. H.
Socas-Navarro helped with the implementation of the MELANIE code and with the
interpretation of the obtained results. Comments on the manuscript
by B.C. Low and J. Trujillo Bueno are gratefully acknowledged.
\end{acknowledgements}

\end{document}